\documentclass[trackchanges, twocolumn]{aastex701}

\begin{document}

\title{Optical and near-infrared nebular-phase spectroscopy of SN 2024ggi: constraints on the structure of the inner ejecta, progenitor mass, and dust. \footnote{This paper includes data gathered with the 6.5 meter Magellan Telescopes located at Las Campanas Observatory, Chile.}}

\author[0000-0002-9216-7996]{E. Hueichap\'an}
\affiliation{Instituto de Estudios Astrof\'isicos, Facultad de Ingenier\'ia y Ciencias, Universidad Diego Portales, Avenida Ejercito Libertador 441, Santiago, Chile}
\affiliation{Millennium Institute of Astrophysics MAS, Nuncio Monseñor Sotero Sanz 100, Off. 104, Providencia, Santiago, Chile}
\email{emilio.hueichapan@mail.udp.cl}

\author[0000-0003-4553-4033]{R\'egis Cartier}
\affiliation{Centro de Astronom\'ia (CITEVA), Universidad de Antofagasta, Avenida Angamos 601, Antofagasta, Chile}
\email{}

\author[0000-0003-1072-2712]{Jose L. Prieto}
\affiliation{Instituto de Estudios Astrof\'isicos, Facultad de Ingenier\'ia y Ciencias, Universidad Diego Portales, Avenida Ejercito Libertador 441, Santiago, Chile}
\affiliation{Millennium Institute of Astrophysics MAS, Nuncio Monseñor Sotero Sanz 100, Off. 104, Providencia, Santiago, Chile}
\email{}

\author[0000-0001-6293-9062]{Carlos Contreras}
\affiliation{Las Campanas Observatory, Carnegie Observatories, Casilla 601, La Serena, Chile}
\email{}

\author[0000-0001-7101-9831]{Aleksandar Cikota}
\affiliation{Gemini Observatory, NSF's National Optical-Infrared Astronomy Research Laboratory, Casilla 603, La Serena, Chile}
\email{}

\author[0000-0001-6540-0767]{Thallis Pessi}
\affiliation{European Southern Observatory, Alonso de C\'ordova 3107, Casilla 19, Santiago, Chile}
\email{}

\author[0000-0002-8686-8737]{Franz E. Bauer}
\affiliation{Instituto de Alta Investigaci\'on, Universidad de Tarapac\'a, Casilla 7D, Arica, Chile}
\email{}

\author[0000-0003-0006-0188]{Giuliano Pignata}
\affiliation{Instituto de Alta Investigaci\'on, Universidad de Tarapac\'a, Casilla 7D, Arica, Chile}
\email{}

\author[0009-0002-0843-0059]{Camila Cardenas}
\affiliation{Centro de Astronom\'ia (CITEVA), Universidad de Antofagasta, Avenida Angamos 601, Antofagasta, Chile}
\email{}

\author[0009-0000-4635-4945]{Sethulakshmi Vazhayil}
\affiliation{Instituto de Estudios Astrof\'isicos, Facultad de Ingenier\'ia y Ciencias, Universidad Diego Portales, Avenida Ejercito Libertador 441, Santiago, Chile}
\affiliation{Millennium Institute of Astrophysics MAS, Nuncio Monseñor Sotero Sanz 100, Off. 104, Providencia, Santiago, Chile}
\email{}

\begin{abstract}

We present optical and near-infrared (NIR) spectroscopic observations of the nearby Type II supernova SN\,2024ggi from 250 and 581 days after the explosion. Comparing the evolution of the [\ion{O}{1}] at 6300, 6363 \text{\AA} doublet normalized to the continuum with spectral models from the literature, we estimate a progenitor star zero-age main-sequence mass ($M_{\mathrm{ZAMS}}$) of $\approx 14$ M$_\odot$. This value is consistent with $M_{\mathrm{ZAMS}}$ reported in the literature from independent methodologies. The nebular spectra are used to study the structure of the inner ejecta. The broad H$\alpha$ line has a full-width at half maximum (FWHM) of $\simeq 3900$ km s$^{-1}$, with small deviations from a symmetric Gaussian profile centred at zero velocity, and the [\ion{O}{1}] doublet is blue-shifted by $\approx -940$ km s$^{-1}$. In the NIR, the nebular spectra reveal double-peaked emission features of \ion{Mg}{1} and [\ion{Fe}{2}] lines between +250 and +319 days, suggesting a bipolar distribution of intermediate mass and iron peak elements in the line-of-sight. Such a double-peaked feature in these NIR lines has not been previously reported. No corresponding asymmetries are observed in the hydrogen lines, suggesting that the asymmetry is mostly confined to intermediate mass and iron peak elements in the innermost core of the supernova ejecta. Additionally, we detect first-overtone carbon monoxide (CO) emission at 2.3,$\mu$m between 250 and 319 days, and a blueshift in the emission lines of H$\alpha$, [\ion{O}{1}], \ion{Mg}{1}], and [\ion{Fe}{2}] in the +581 day optical spectrum, consistent with dust formation in the ejecta.

\end{abstract}



\section{Introduction}

Transient surveys scrutinize the night sky with an unprecedented high cadence and depth, finding young supernovae (SNe) caught within hours of their explosion. Two recent examples of young discoveries and rapid spectroscopy are Type II SN\,2023ixf and SN\,2024ggi. Both exploded in nearby galaxies at distances of $\sim 7$ Mpc, were discovered within a day of their explosion, and were classified spectroscopically within a day of their discovery. High-quality, multi-wavelength (UV through near-IR) photometric observations and time series spectra were obtained for these groundbreaking events, used to inform new modelling techniques, and provide tight constraints on pre-SN explosion scenarios \citep{bersten24, ertini25, kozyreva25, dessart25}. Their progenitor stars were detected in HST and Spitzer pre-explosion images \citep{kilpatrick23,jencson23,pledger23,niu23A,vandyk24,qin24,xiang24}, and their early spectra reveal the final moments of their progenitor stars and the extension and density of the dense circumstellar medium (CSM) surrounding them \citep{jacobsongalan23,jacobsongalan24b}.

\begin{deluxetable*}{lcccccc}[ht!]
\tablecaption{Summary of spectroscopic observations of SN~2024ggi. The phase is measured relative to the time of first light estimated by \cite{Pessi24}, MJD = 60410.89 $\pm$ 0.14. }\label{tab:spec_log}
\tablehead{
\colhead{Date (UT)} & \colhead{MJD} & \colhead{Phase (days)} &
\colhead{Instrument/Telescope} & \colhead{Wavelength Range (\AA)} &
\colhead{Dispersion (\AA/pix)} & \colhead{Resolution (FWHM; \AA)}
}
\startdata
2024-12-17 & 60661.3 & 249.8 & Flamingos-2/Gemini-S    & 8750--17,000     & 6.0 & 18 \\
2024-12-18 & 60662.3 & 250.8 & Flamingos-2/Gemini-S    & 13,300--24,700   & 7.6 & 22 \\
2025-01-05 & 60680.4 & 268.8 & GMOS-S/Gemini-S         & 5400--9500       & 1.3 & 5.5 \\
2025-01-13 & 60688.3 & 276.7 & Flamingos-2/Gemini-S    & 13,300--24,700   & 7.6 & 22 \\
2025-01-13 & 60688.3 & 276.8 & Flamingos-2/Gemini-S    & 8750--17,000     & 6.0 & 18 \\
2025-02-24 & 60730.2 & 318.5 & Goodman/SOAR            & 3600--9000       & 2.0 & 5.5 \\
2025-02-24 & 60730.2 & 318.6 & TripleSpec/SOAR         & 9400--24,650     & 2.2 & 4.5 \\
2025-05-02 & 60797.1 & 385.3 & Goodman/SOAR            & 3600--9000       & 2.0 & 5.5 \\
2025-05-13 * & 60808.1 & 396.3 & LDSS-3/Magellan Clay    & 3800--10,300   & 2.0 & 8.0 \\
2025-06-06 & 60832.0 & 420.2 & LDSS-3/Magellan Clay    & 3800--10,300     & 2.0 & 8.0 \\
2025-07-14 & 60871.0 & 459.1 & LDSS-3/Magellan Clay    & 3800--10,300     & 2.0 & 6.7 \\
2025-11-13 & 60993.3 & 581.1 & LDSS-3/Magellan Clay    & 3800--10,300     & 2.0 & 9.6 \\
\enddata
\tablecomments{(*): The spectrum shown on May 13th, 2025, is part of the averaged spectrum shown in \cite{ferrari25}.}
\end{deluxetable*}

SN\,2024ggi is a type II SN discovered only 6\,hrs after the time of first light \citep[see e.g.,][]{Pessi24} by the ATLAS survey \citep{tonry24}, located in the nearby galaxy NGC 3621 at $7.24 \pm 0.20$\,Mpc \citep{saha06}. At early times the spectra of SN\,2024ggi showed narrow lines from unshocked CSM located in a confined region in the vicinity of the progenitor star \citep{jacobsongalan24b,Pessi24,shrestha24, zhang24}. Two early-time high-resolution spectra the night after discovery, at 26.6\,hrs and 33.8\,hrs after the first light, revealed that the CSM suffers radiative acceleration, and the \ion{He}{1} CSM lines disappear by 1.4\,days after the explosion, implying that the helium in the CSM is completely photoionized at \ion{He}{2} \citep{Pessi24}.

The progenitor star of SN\,2024ggi is detected in pre-explosion images of ground-based telescopes \citep{perezfournon24, chen25} and space telescopes such as {\it Spitzer} and {\it Hubble} \citep{xiang24}, yielding an estimated $M_{ZAMS} = 13 \pm 1$~ M$_{\odot}$ \citep{xiang24}. \citet{ertini25} presented  hydrodynamical modeling of the bolometric light curve of SN\,2024ggi finding a progenitor mass of $M_{\mathrm{ZAMS}} = 15$~M$_{\odot}$, with a pre-SN mass and radius of $14.1$~M$_{\odot}$ and 517~R$_{\odot}$, respectively, an explosion energy of $1.2 \times 10^{51}$~erg, and a synthetized $^{56}$Ni mass below $0.035$~M$_{\odot}$. Recently, \cite{dessart25} presented optical and near/mid-IR nebular-phase spectra ($\sim 275-400$~days after explosion) of the SN using ground-based telescopes and {\it JWST}. They use radiative transfer modelling to put constraints on the progenitor and explosion, finding a progenitor mass of $M_{\mathrm{ZAMS}} = 15.2$~M$_{\odot}$, an explosion energy of $\sim 1.0 \times 10^{51}$~erg and a $^{56}$Ni mass of $0.06$~M$_{\odot}$. Finally, \cite{ferrari25} estimates a range of masses between $12 - 15$ M$_\odot$ by matching the nebular-phase spectra with spectral models, and a slightly lower mass range of $10 - 12$ by measuring the [\ion{O}{1}]/[\ion{Ca}{2}] flux ratio.

 In this paper, we present optical and near-infrared nebular-phase spectra of SN\,2024ggi. After the hydrogen envelope recombines, the ejecta opacity of SNe II drops significantly, entering into the nebular phase, thus enabling us to probe the innermost layers of the supernova ejecta. The view of the inner ejecta can be used to study the velocity distribution and the abundance of different elements, thus constraining the mass of the progenitor star \citep{jerkstrand14,dessart21,fang25} and the structure of the inner ejecta \citep{leonard06,bose19,leonard21, vasylyev25}. In addition, the view of the inner ejecta also enables the study of molecules such as carbon monoxide (CO) and dust formed in the SN ejecta \citep{rho18,tinyanont19,rho21,shahbandeh23,stritzinger24,cartier24,dessart25,park25,pearson25}.

In this paper, we adopt the date of first light of SN~2024ggi measured by \cite{Pessi24} of $t_0=60410.89 \pm 0.14$ (MJD) and a total reddening in the line of sight, including Galactic and host galaxy reddening, of $E(B-V) = 0.16$~mag.

\begin{figure*}[ht!]
\centering
\includegraphics[width=\linewidth]{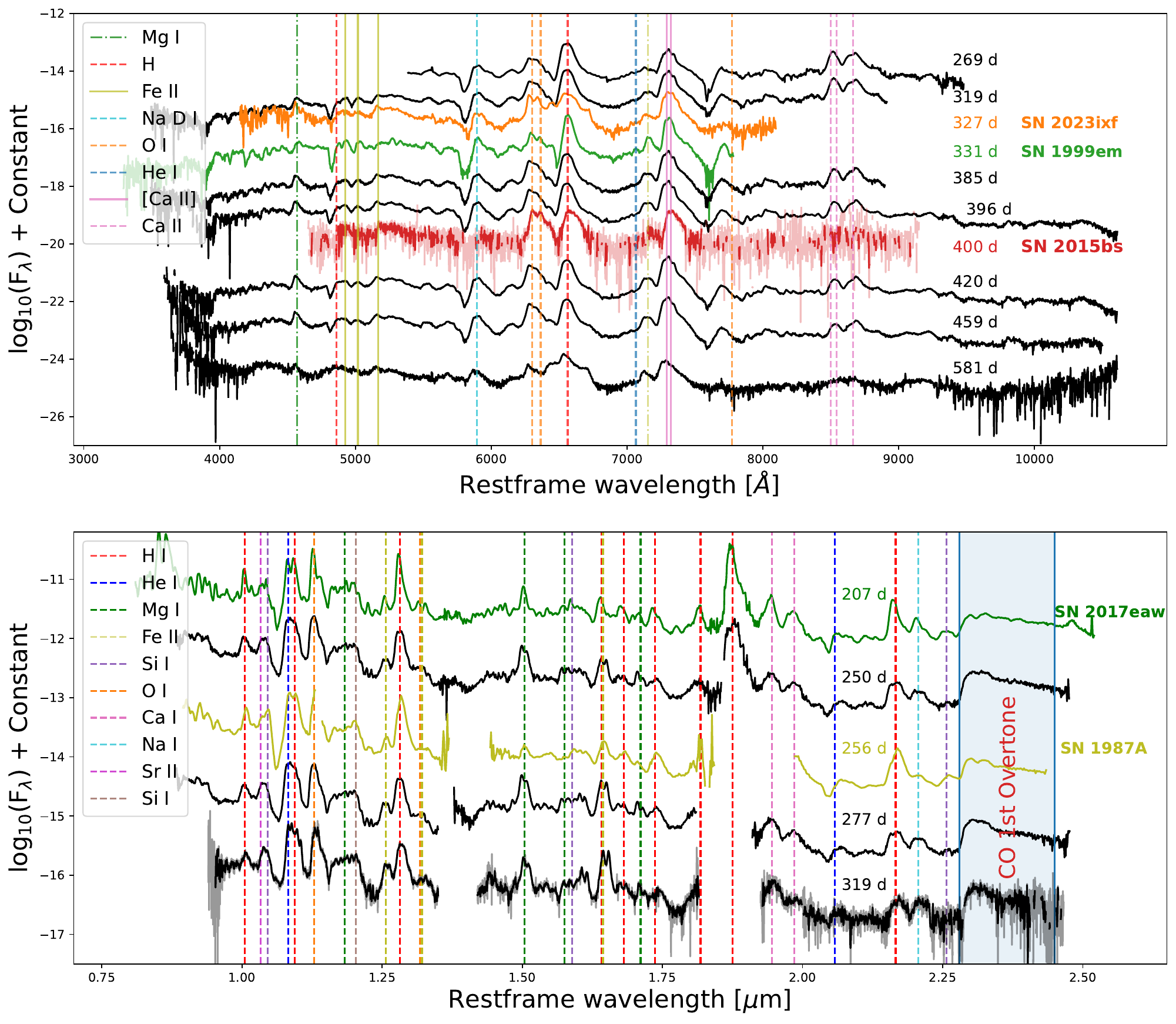}
\caption{Nebular-phase spectral time series of SN 2024ggi shown in black. Upper panel: Optical spectra from 269 to 420 days post-explosion, with SN 1999em \citep{Leonard02}, SN 2015bs \citep{anderson18}, and SN 2023ixf \citep{Michel25} shown for comparison. The 400d spectrum of SN 2015bs is smoothed with a Savitzky–Golay filter. Lower panel: Near-IR spectra from 250 to 319 days. The 319d spectrum of SN 2024ggi is smoothed. NIR spectra of SN 1987A \citep{Bouchet91} and SN 2017eaw \citep{rho18} are included for comparison. Vertical lines mark rest-frame wavelengths of key identified ions.}
\label{fig:time-series}
\end{figure*}

\section{Data}

Optical and near-infrared (NIR) spectra of SN\,2024ggi were obtained with Flamingos-2 \citep{eikenberry04S, eikenberry12} and GMOS-S \citep{hook04, gimeno16} spectrographs mounted on the Gemini-South 8-m telescope, Goodman \citep{clemens04} and Triple-Spec \citep{schlawin14} spectrographs on the SOAR telescope, and with the LDSS-3 spectrograph mounted on the Magellan Clay telescope. Standard reduction steps were used for data reduction \citep[see][for details]{cartier24}. These steps include the basic processing of the 2D frames, wavelength calibration, spectral extraction, and flux calibration. The flux calibration and telluric correction of the NIR spectra was performed with the {\sc xtellcorr} task \citep{vacca03}, using an A0 star observed close in time and airmass to the SN.

A spectroscopic log of the optical and near-IR observations is presented in Table~\ref{tab:spec_log}. The spectra are presented in Figure \ref{fig:time-series}.

\section{Analysis}

\subsection{Spectral line identification}

In Figure~\ref{fig:time-series}, we present seven optical and three near-infrared nebular-phase spectra of SN~2024ggi, spanning from 250 to 581 days after the explosion. The optical spectra are characterized by strong H$\alpha$ in emission, a broad emission profile of the \text{\ion{Na}{1\,D}} 5890, 5896 \text{\AA} doublet, and the presence of [\ion{Fe}{2}] 7155 \text{\AA}. The [\ion{O}{1}] 6300, 6363 \text{\AA} doublet is fully blended, producing a boxy profile. Each peak of the doublet is blue-shifted by approximately $-940$~km~s$^{-1}$, maintaining the expected 63\,\text{\AA} separation between the components. This profile is well reproduced by two Gaussians with FWHM $\sim 2830$~km~s$^{-1}$, centred at 6280 and 6343 \text{\AA}. This profile differs from other supernovae shown in Figure \ref{fig:time-series}, where the two peaks of the [\ion{O}{1}] doublet are clearly distinguished. In the blue part of the spectra, we observe a strong and blueshifted emission of \ion{Mg}{1} 4751\text{\AA}, weak H$\beta$, and \ion{Fe}{2} at 4924, 5018, 5169 \text{\AA} lines. At longer wavelengths, we observe a strong emission of [\ion{Ca}{2}] at 7291, 7324 \text{\AA} doublet, and \ion{Ca}{2} at 8498, 8542, 8662 \text{\AA}. 

\begin{figure}[ht!]
    \centering
    \includegraphics[width=1.1\linewidth]{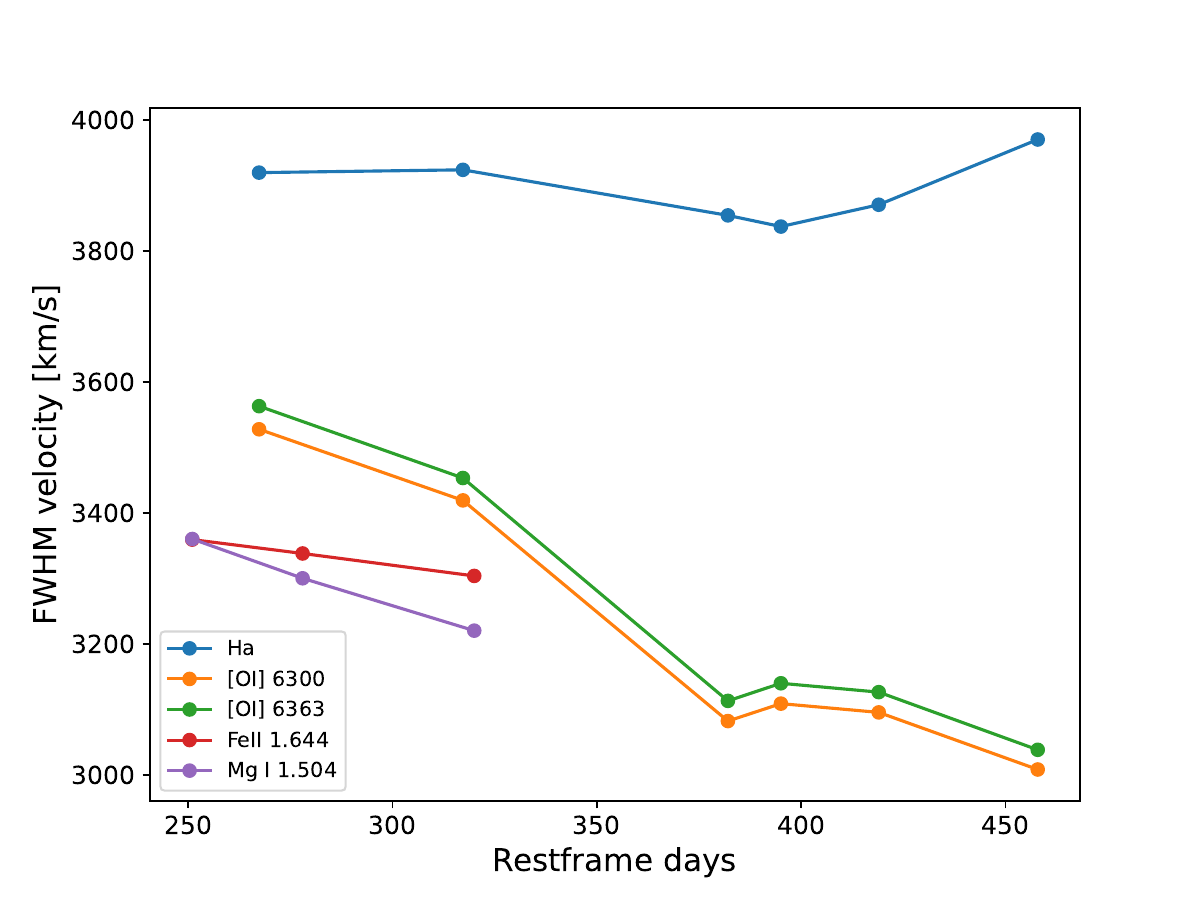}
    \caption{FWHM velocities from single Gaussian fits to the H$\alpha$, [\ion{O}{1}] 6300, 6363 \text{\AA} doublet, \ion{Mg}{1} 1.504$~\mu$m, and [\ion{Fe}{2}] 1.644$~\mu$m line profiles.}
    \label{fig:fwhm}
\end{figure}

In the NIR spectra we observe \ion{H}{1} emission lines of P$\delta$, P$\gamma$, P$\beta$, Br$\gamma$, emission lines of intermediate-mass elements \ion{O}{1}, \ion{Ca}{1}, \ion{Na}{1 D}, and \ion{Mg}{1}, and iron-peak elements \ion{Si}{1} and [\ion{Fe}{2}]. The P$\beta$ and P$\gamma$ lines are broad compared with the line profiles of SN\,1987A and SN\,2017eaw (see Figure~\ref{fig:time-series}). The P$\gamma$ line is blended in a single broad spectral feature with \ion{He}{1} $ 1.083$ $\mu$m, and the Br$\gamma$ line shows a broad, boxy profile. The \ion{Sr}{2} $1.033\, \mu m$ and \ion{Si}{1} $1.046\, \mu m$ lines are also blended in a single broad feature. 
In Section \ref{sec:Fe_Mg} we discuss the double-peak profiles of the [\ion{Fe}{2}] and \ion{Mg}{1} lines. Finally, the spectral signature of the first CO overtone around 2.3 $\mu$m is detected in the three NIR spectra from 250 to 319\,days (see Figure \ref{fig:time-series}). 
The FWHM velocities of the H$\alpha$ and [\ion{O}{1}] doublet lines in the optical, and of the \ion{Mg}{1} and [\ion{Fe}{2}] lines in the NIR were calculated by fitting a Gaussian profile to each line profile. The largest FWHM correspond to the H$\alpha$ profile, with a velocity decreasing from $3920$ to $3870$\,km\,s$^{-1}$ (Figure \ref{fig:fwhm}) between 269 d and 420 d. In turn, each component of the [\ion{O}{1}] doublet has a FWHM of $\sim 3550$\,km\,s$^{-1}$ at 250\,days that decreases to $\sim 3100$\,km\,s$^{-1}$ at 319\,d (see Figure~\ref{fig:fwhm}).
The \ion{Mg}{1} $1.504\, \mu m$ and [\ion{Fe}{2}] $1.644\, \mu m$ lines were fitted with a single Gaussian and with two Gaussians to model the double-peaked profiles. The difference between the FWHM derived from the single-Gaussian and the combined fit is negligible, so we adopt the FWHM velocity from the single-Gaussian fit, obtaining average velocities of $\sim $3330\,km\,s$^{-1}$ for [\ion{Fe}{2}]. The same procedure applied to \ion{Mg}{1} yields a FWHM velocity decline from 3360 to 3220\,km\,s$^{-1}$. In the NIR spectra, the [\ion{Fe}{2}] 1.257\,$\mu$m, 1.644\,$\mu$m and \ion{Mg}{1} 1.504\,$\mu$m, 1.711\,$\mu$m lines exhibit double-peaked emission profiles. The FWHM velocity evolution of these profiles is shown in Figure \ref{fig:fwhm}.

\subsection{\ion{Fe}{2} and \ion{Mg}{1} profiles in the near-IR spectra}  \label{sec:Fe_Mg}

\begin{figure*}[ht]
    \centering
    \includegraphics[width=\linewidth]{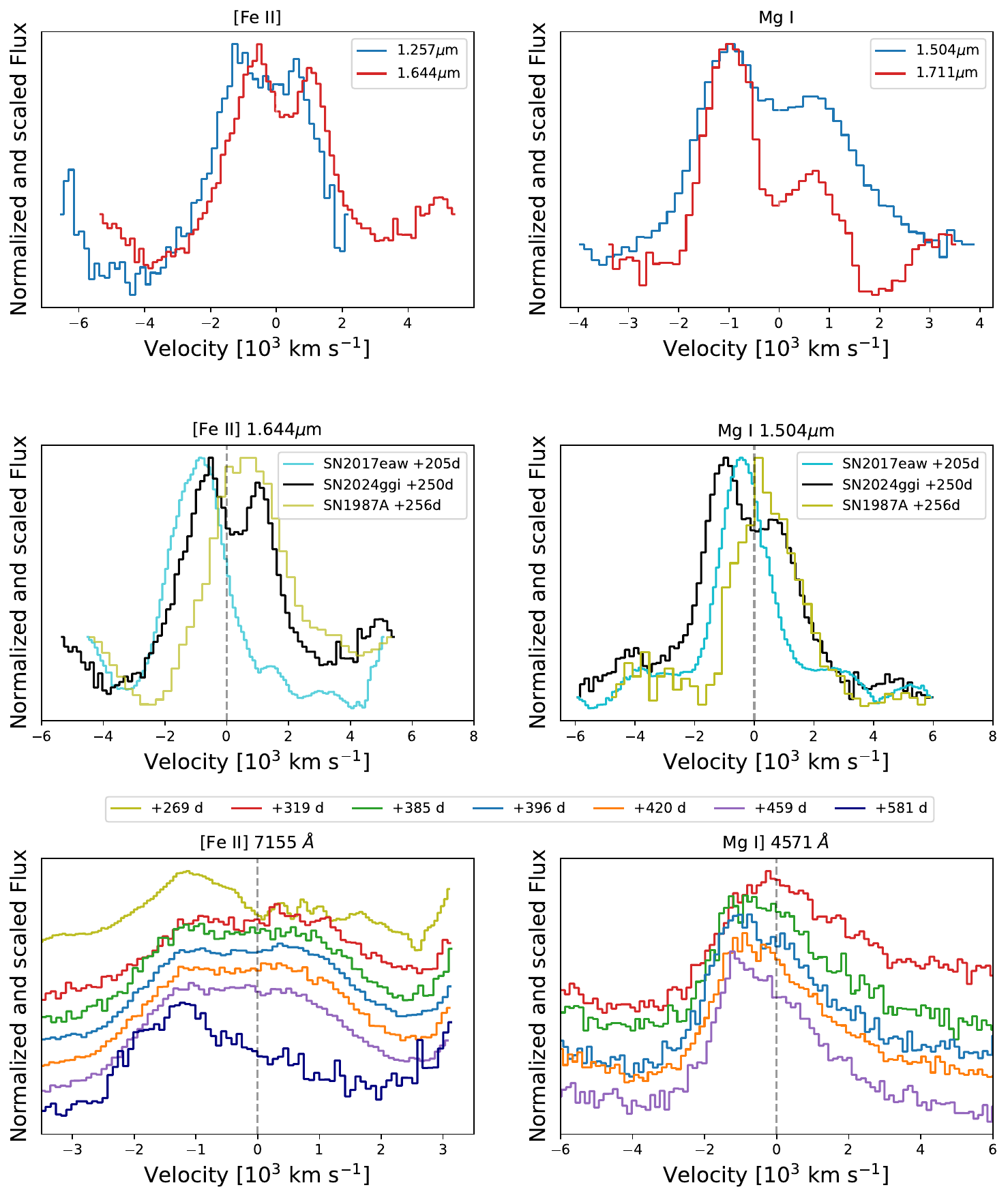}
    \caption{Near-IR and optical [\ion{Fe}{2}] and \ion{Mg}{1} line profiles comparison. Upper left: Comparison between [\ion{Fe}{2}] 1.257 $\mu$m and [\ion{Fe}{2}] 1.664~ $\mu$m  emission at 250 days. Upper right: Comparison between \ion{Mg}{1} 1.504 $\mu$m and 1.711$ \mu$m emission at 250d. In both panels, a double-peaked profile is observed. Middle left: [\ion{Fe}{2}] 1.644 $\mu$m from SN 2024ggi at 250d (black) compared with SN 1987A \citep{Bouchet91}, and SN 2017eaw \citep{rho18}. Middle right: \ion{Mg}{1} 1.504 $\mu$m from SN 2024ggi, compared with SN 1987A \citep{Bouchet91}, and SN 2017eaw \citep{rho18}. Bottom left: [\ion{Fe}{2}] 7155 \text{\AA} emission profile. The emission displays a broad and flat profile. Bottom right: \ion{Mg}{1}] 4571 \text{\AA} emission profile. At 385 days post first light, a blueshift of $\sim $ 1.300 km s $^{-1}$ is observed.}
    \label{fig:iron-comp}
\end{figure*}

The [\ion{Fe}{2}] and \ion{Mg}{1} features exhibit double-peaked emission lines in the NIR spectra. The top panels of Figure \ref{fig:iron-comp} show a comparison of the [\ion{Fe}{2}] $1.257$ $\mu$m and $1.664$ $\mu$m lines, and of the \ion{Mg}{1} $1.540$ $\mu$m and $1.711$ $\mu$m of SN\,2024ggi. The peaks of the \ion{Mg}{1} lines show an excellent match, giving us confidence in the double-peaked nature of this emission line. The correspondence of double-peak positions of the [\ion{Fe}{2}] lines is good, but they could be affected by other ions. In the bottom panel of Figure \ref{fig:iron-comp} we compare the [\ion{Fe}{2}] $1.664$ $\mu$m and \ion{Mg}{1} $1.540$ $\mu$m profiles of SN\,20224ggi with the profiles of SN~2017eaw \citep{rho18} and SN~1987A\citep{Bouchet91}. The double-peaked emission of SN\,2024ggi is distinct from any other spectra found in the literature. 

Each feature shows a blue and a red peak relative to the rest wavelength, with different velocities, and there is little to no evolution between the three epochs. An unexpected double-peaked profile of the single emission line of \ion{Mg}{1} is clearly observable at  $1.504$ $\mu$m and $1.771$ $\mu$m. Both red and blue peaks are consistent among themselves, with velocities of $-993$ km s$^{-1}$ and \mbox{$+720$ km s$^{-1}$} for the blue and red peaks, respectively. The [\ion{Fe}{2}] $1.257$ $\mu$m emission line shows a similar profile with a blue peak at $-1280$ km s$^{-1}$ and a red peak at $+604$ km s$^{-1}$. The same ion, at $1.664$ $\mu$m, exhibits similar peaks but with different velocities. We measure a blue peak at $-532$ km s$^{-1}$ and a red peak at $+933$ km s$^{-1}$. 

Even though we do not observe double-peaked emission profiles of [\ion{Fe}{2}] and \ion{Mg}{1}] in the optical spectra, we do note that [\ion{Fe}{2}] 7155 \text{\AA} displays a broad profile with peaks at $-869$ km s$^{-1}$ and $+1215$ km s$^{-1}$ On the other hand, the \ion{Mg}{1}] 4571 \text{\AA} emission feature is blueshifted by $-1311$ km s$^{-1}$. Both values are consistent with what we observed in the NIR spectra. 

We compared the [\ion{Fe}{2}] $1.644$ $\mu$m with spectra found in literature at similar phases. The  [\ion{Fe}{2}] $1.644$ $\mu$m in SN\,2017eaw \citep{rho18} is blue-shifted, with the peak emission consistent with the blue peak in SN\,2024ggi. On the other hand, the red peak is consistent with the peak of the line observed in SN 1987A \citep{Bouchet91}.

\subsection{CO 1st overtone and dust formation}

Carbon monoxide plays a key role in cooling the ejecta and is among the earliest molecules to form in CCSNe. A clear detection of the first CO overtone at $\sim$2.3\,$\mu$m is observed in three NIR spectra obtained between +250 and +319 days. Although the CO first-overtone emission is not perfectly smooth, it lacks clear band-head substructures, which may indicate a relatively cool and/or clumpy gas with a high rotational population \citep{Spyromilio90}. As one of the first molecules to form in CCSNe, CO provides efficient radiative cooling, which is known to enhance dust formation \citep{LiMcCray92}.

\begin{figure}[ht]
    \centering
    \includegraphics[width=\linewidth]{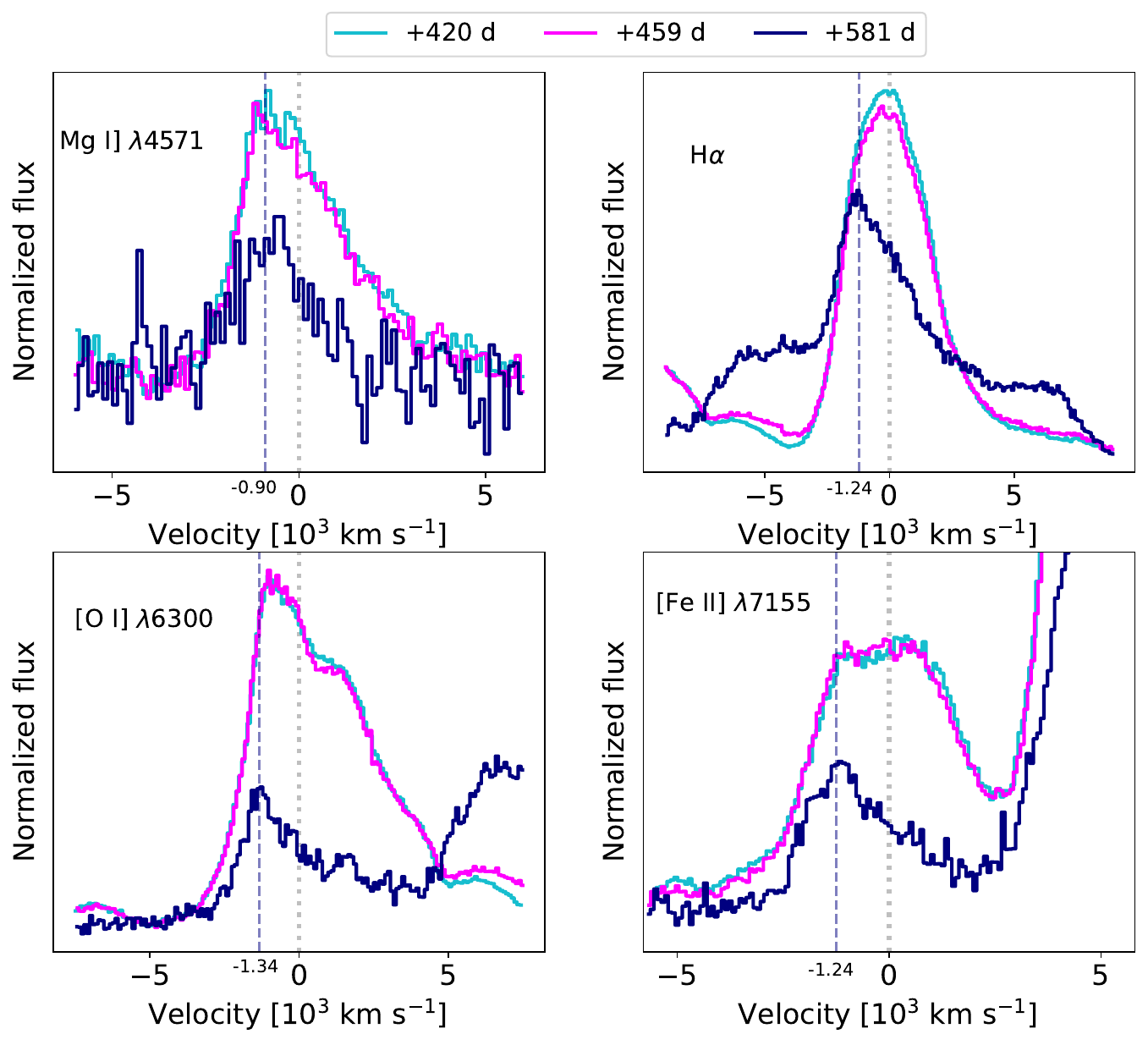}
    \caption{Line profile of the emission lines affected for dust formation. The grey dotted line denotes zero velocity at redshift $z$ = 0.002435. The black dashed line marks the peak of the emission of each ion.}
    \label{fig:dust_effect}
\end{figure}

The optical spectra obtained at +420 and +459 days show no significant evolution in the emission-line profiles. As shown in Figure~\ref{fig:dust_effect}, the lines of H${\alpha}$, \ion{Mg}{1}], [\ion{O}{1}], and [\ion{Fe}{2}] remain largely symmetric over this period.

Clear changes only become apparent in the last optical spectrum, at +581 days. At this epoch, all these lines exhibit a consistent blueshift, ranging from approximately $-1300$ to $-900$ km s$^{-1}$. The emission profiles of both [\ion{O}{1}] and [\ion{Fe}{2}] have lost most of their redshifted emission, transitioning from an initially flat morphology to predominantly blueshifted profiles, with velocity centroids of $-1340$ and $-1240$ km s$^{-1}$, respectively. In this scenario, the observed blueshift arises because the newly formed dust absorbs and scatters photons emitted from the receding side of the ejecta, suppressing the red wing of the line profile.

\begin{figure}[ht]
    \centering
    \includegraphics[width=\linewidth]{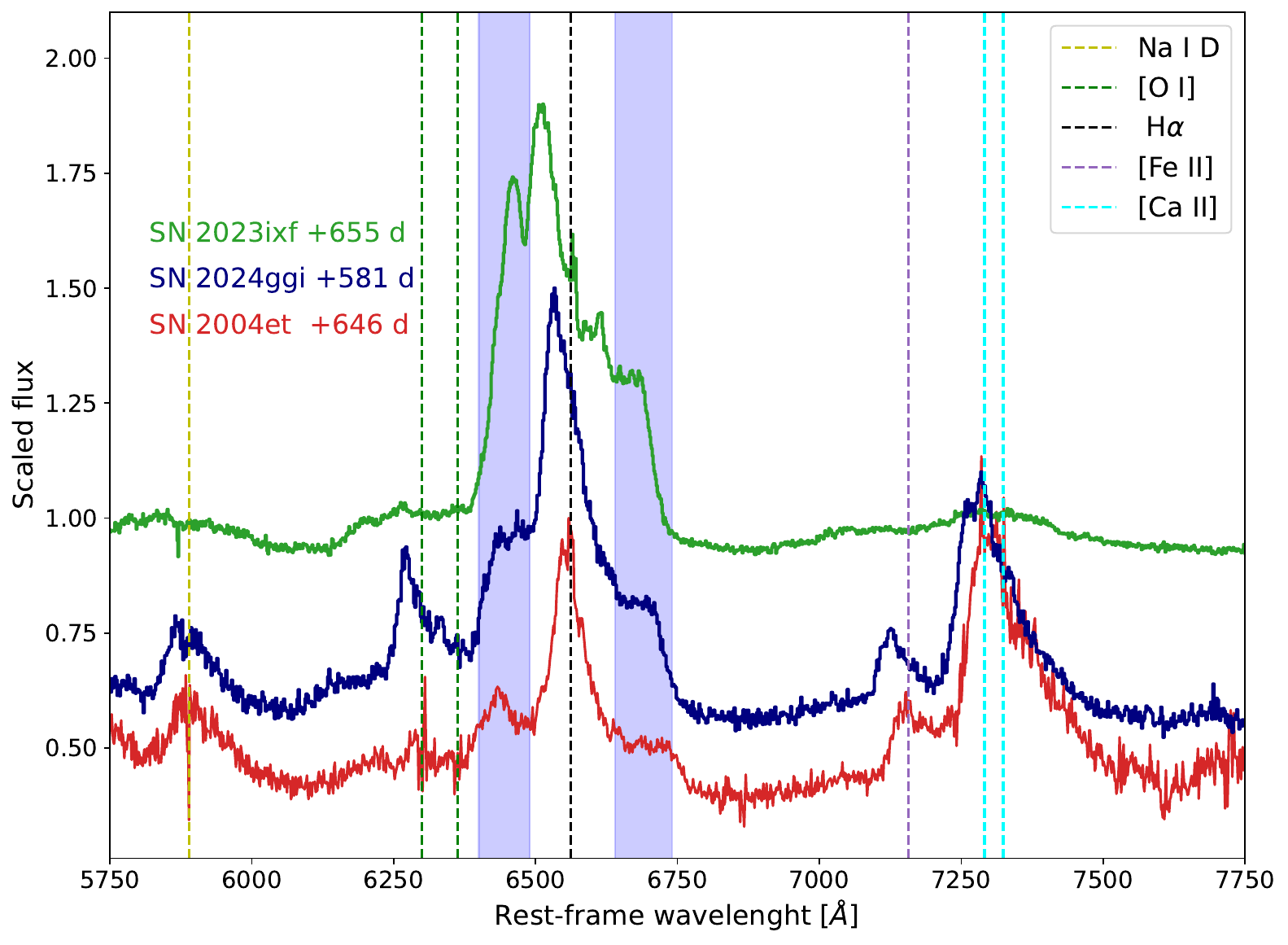}
    \caption{Comparison between nebular spectra of SN 2024ggi, SN 2023ixf \citep{Jacobsongalan25}, and SN 2004et \citep{Fabbri11}. Broad features in the red and blue sections of the H$\alpha$ emission profile are observed}.
    \label{fig:24ggi04et}
\end{figure}

The H$\alpha$ emission in the later spectra exhibits a blueshift of $-1240$ km s$^{-1}$ and two prominent features towards the red and blue parts of the emission (see Figure \ref{fig:24ggi04et}). When comparing the blueshift of the peak of H$\alpha$ is $-1240$ km s$^{-1}$, with other supernovae, it is two times larger than the blueshift of $\sim$ $-600$ km s$^{-1}$ measured for SN 2004et at 640 days post explosion, and $-2000$ km s$^{-1}$ less than the shift observed in SN 2023ixf, but this emission is in part, explained by ejecta-CSM interaction than dust formation \citep{Jacobsongalan25}. Similar red and blue structures were reported in the emission of H$\alpha$ of SN 2004et and mentioned by \cite{Fabbri11}, and \cite{Kotak09} showed how those features evolve into a broad boxy H$\alpha$ component. Later, \cite{NiculescuDuvaz22} uses \texttt{DAMOCLES} to model the emission line, obtaining the same features by assuming 100\% of clumping and estimating a mass of $\approx$ 5.00 $\times$ 10$^{-4}$ M$_\odot$.

\subsection{Constraints on the Progenitor Mass}

To estimate the $M_{ZAMS}$ mass of the progenitor star of SN~2024ggi from the [\ion{O}{1}] 6300, 6363 \text{\AA} doublet, we followed the procedure described in \citet{fang25} using the optical nebular-phase spectra. We normalized the de-reddened optical spectra using the integral between $5000\,\text{\AA}$ and $8500\,\text{\AA}$ regions.
To define a local continuum to the [\ion{O}{1}] doublet and H$\alpha$ lines, we performed a linear fit to the regions [$6020, 6090$] \text{\AA} and [$6800, 6850$] \text{\AA}, and subtracted this local continuum from the spectra. 
We then fit three Gaussian profiles: two for the [\ion{O}{1}] $6300, 6363$ \text{\AA} doublet and one for H$\alpha$ profile, along with an additional Gaussian for an extra component observed in some SNe at approximately $6430$ \text{\AA}. For the [\ion{O}{1}] doublet, we fixed the separation between the central wavelengths of the Gaussians at 63\, \text{\AA} and constrained $\sigma$ of both Gaussians to have the same value, as they correspond to the same ion. Only these two forbidden [\ion{O}{1}] transitions are expected to dominate this wavelength region, and no additional lines are known to significantly contribute at these epochs (See \cite{LiMcCray92, Chugai92, Taubenberger09}). Later models have shown that making the two-component Gaussian fitting a standard treatment of the [\ion{O}{1}] doublet. 

We finally calculated the flux ratio $f_{[\text{\ion{O}{1}}],\mathrm{reg}}$, 
defined as: $f_{[\text{O I}], \mathrm{reg}} = \frac{[\text{O I}]}{1 - H\alpha}$, where $f_{\text{O I}}$ and {H$\alpha$} are the total fluxes calculated from the integral of the Gaussian fits of each line. We also directly integrated the total fluxes of the [\ion{O}{1}] feature between 6220\,\text{\AA} and 6400\,\text{\AA}, and H$\alpha$ between 6420\,\text{\AA} and 6800\,\text{\AA}. The ratio calculated using the direct integral method gives slightly higher values for all spectra than using the Gaussian fits. However, the difference is not significant to our conclusions. 

To estimate statistical uncertainties, we randomly varied the edges of the two continuum regions within a window corresponding to 10\% of the length of each region, producing several sets of Gaussian fits. We used the mean of the fluxes to calculate $f_{[\text{\ion{O}{1}}], \mathrm{reg}}$, and the standard deviation as the associated uncertainty. As each spectrum has a high signal-to-noise ratio, the variations in the continuum regions had little impact, and the resulting uncertainty was approximately $\sim1\%$.

\begin{figure*}
    \centering
    \includegraphics[width=\linewidth]{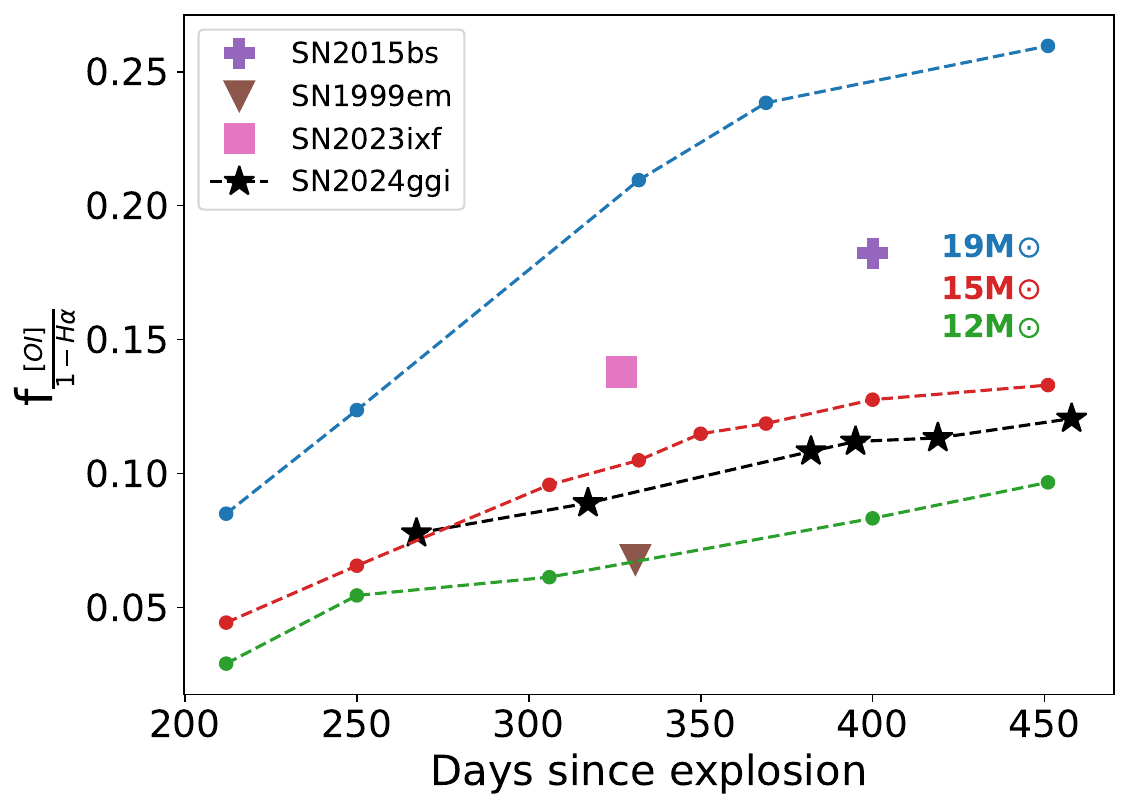 }
    \caption{Time evolution of the fractional flux of [\ion{O}{1}] 6300, 6363 \text{\AA} compared to the continuum from the nebular spectra of SN 2024ggi compared with the results of the models from \cite{jerkstrand14}. We also include the measurements obtained from the nebular-phase optical spectra of the Type II SN 1999em \citep{Leonard02}, SN 2015bs \citep{anderson18}, and SN 2023ixf \citep{Michel25}.}
    \label{fig:progenitor}
\end{figure*}

We repeat the same Gaussian fitting procedure with the nebular-phase spectral models of Type II SNe presented in \cite{jerkstrand14} for different progenitor masses: M$_{\rm ZAMS}$=12, 15, and 19~M$_{\odot}$, and compare the results with the line ratios from SN~2024ggi. We obtain an estimated ZAMS mass of the progenitor star of $\approx 14$~M$_{\odot}$ after interpolating linearly the ratios from the models at different epochs. We present the temporal evolution of the line ratios of SN~2024ggi and the models from \citep{jerkstrand14} in Figure~\ref{fig:progenitor}. In the figure, we also include results from published nebular-phase optical spectra of SN\,1999em \citep{Leonard02}, SN 2015bs \citep{anderson18}, and SN\,2023ixf \citep{Michel25}.

\section{Discussion and conclusions}

We present a set of late-time optical and near-infrared (NIR) nebular-phase spectra of the nearby Type II SN 2024ggi, spanning a time range of 250 to 420 days after explosion. The NIR spectra of SN\,2024ggi presented here constitute one of the best-observed NIR spectral sequences at nebular phase, alongside SN\,1987A \citep{Bouchet91}, SN\,2017eaw \citep{rho18, tinyanont19}, and SN\,2023ixf \citep{park25}.

We observe double-peaked profiles of the emission lines of \ion{Mg}{1} and [\ion{Fe}{2}] in the NIR spectra. As mentioned in Section \ref{sec:Fe_Mg}, we do not observe a clear double-peaked emission in the [\ion{Fe}{2}] in the optical spectra, but it is observable in the spectra at 287d reported by \cite{ferrari25}. We fitted 2 Gaussians to the emission at 7155\,\text{\AA} in the earliest optical spectrum we have at 269 d and found that the two components are consistent with the double peak observed in the NIR spectra. The two broad components have FWHM velocities of 1977 km s$^{-1}$ and 1965 km s$^{-1}$ for the red and blue components, respectively. We also fit a single Gaussian to the emission and obtained a FWHM velocity of 3447 km s$^{-1}$, consistent with the values obtained in Section \ref{sec:Fe_Mg}. \cite{dessart25} presents nebular spectra at similar epochs of SN\,2024ggi obtained with JWST, showing a clear double-peaked emission of [\ion{Ni}{1}] at 3.119 $\mu$m with velocities that are consistent with those we measure here for \ion{Mg}{1} and [\ion{Fe}{2}].

Although several Type II SNe have shown double-peaked profiles in H$\alpha$ \textup{(SN 2004dj, \citealt{chugai06}; ASASSN-16at, \citealt{bose19})}, this feature has not been reported in the NIR. While this double-peaked profile can be the consequence of different scenarios, we interpret this as a strong asymmetry in the inner core of the ejecta. There seems to be no clear asymmetry in the H$\alpha$, H$\beta$, or the Paschen series profiles. We find it highly unlikely for this to be produced by line overlapping, as there are four different double-peaked emission lines for two different elements in our observations. This becomes evident after checking the [\ion{Ni}{1}] 3.119 $\mu$m feature, a line that is virtually isolated with a double-peaked profile \citep{dessart25}. We also discard the idea that these profiles are the result of dust formation. It is well-known that newly formed dust particles can produce asymmetries in the emission lines, particularly towards the red part of the emission \citep{Bevan2016}, but in simulations, dust does not produce the double-peaked emission features such as the ones we see in SN 2024ggi. Also, none of the hydrogen lines shows the blueshift expected for a dust-affected line \citep{Bevan2016} and observed in SN\, 2017eaw \citep{rho18}.

This absence, together with the boxy, blueshifted, and/or asymmetric profiles observed only in the intermediate-mass and iron-group elements, indicates that the asymmetry should have happened in the inner core of the ejecta. For example, a bipolar $^{56}$Ni distribution or clumps moving in opposite directions can explain these structures in the spectra. Asymmetries in the inner core of the ejecta of Type II core-collapse SNe have been identified through different observational techniques that are broadly consistent with 3D neutrino-driven explosions \citep[e.g.,][]{wang2002,Milisavljevic13,sinnot13,grefenstette14,Wongwathanarat17,hollandashford20,orlando21,jerkstrand25,vasylyev25}. 

\begin{figure}
    \centering
    \includegraphics[width=\linewidth]{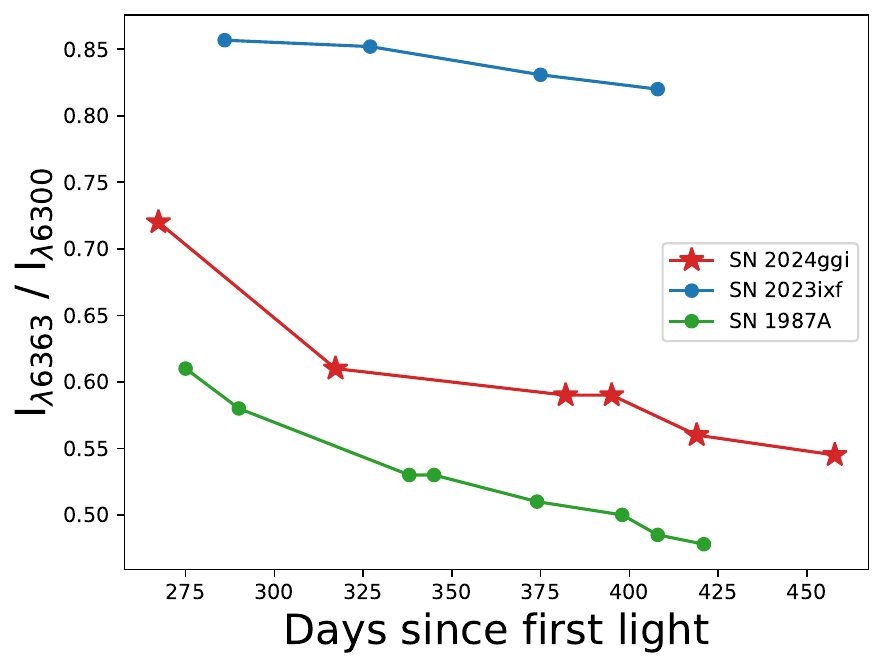}
    \caption{Evolution of the intensity ratio between [\ion{O}{1}] 6363 \text{\AA} and [\ion{O}{1}] 6300 \text{\AA} for SN\,1987A \citep{Phillips90}, SN\,2023ixf \citep{Michel25}, and SN\,2024ggi (This work).}
    \label{fig:IrIb}
\end{figure}

We estimate a flux ratio of [\ion{O}{1}] 6363 \text{\AA} to 6300 \text{\AA} of 0.73 in the +269 days spectrum, which evolves with time to 0.58 in the +420 days spectrum, closer to the 1:3 optically thin ratio expected for the nebular times. This discrepancy also supports the asymmetries or clumping scenario, as the thickness of the ejecta can be explained by an overdensity in the line of sight produced by \ion{O}{0} clumps. In Figure \ref{fig:IrIb} we show that, although these values are higher for an optically thin regime expected for these epochs \citep{Spyromilio90}, they are not unique. Deviations from the theoretical values depends on the density structure of the emitting region \citep{Maguire12}. The same ratio for SN 2023ixf \citep{Michel25} is clearly higher and with little evolution, while the evolution of SN\,2024ggi goes towards the optically thin regime. Figure \ref{fig:IrIb} also shows SN\,1987A \citep{Phillips90} for comparison purposes.

A clear emission of the Carbon Monoxide (CO) molecule is detected in the three epochs of NIR nebular-phase spectra around 2.3 $\mu$m. According to the CO first overtone models reported by \cite{park25}, the lack of band-heads in the emission indicates relatively lower temperatures. For SN 2024ggi, we do not observe any clear band-heads; therefore, we estimate a temperature of $\sim 2000 $~K, which corresponds to the threshold at which such strong band-heads begin to appear. This is roughly consistent with the values estimated in \cite{Mera2025} by modelling the spectral properties of the dust. The spectra +581 days post explosion show a consistent blueshift in the H$\alpha$, [\ion{O}{1}], \ion{Mg}{1}], [\ion{Fe}{2}], ranging between $-1340$ and $-900$ km s$^{-1}$. This suppression of the red-side emission in the optical lines, together with the detection of CO provides consistent evidence for dust formation in the ejecta at +581 days. The broad features observed in the red and blue sections of H$\alpha$ might indicate a highly clumped dust \citep{Bevan2016, NiculescuDuvaz22}. 

Using the flux ratios between the [\ion{O}{1}] 6300, 6363 \text{\AA} doublet and the continuum of SN\,2024ggi estimated at different epochs, and the flux ratio of the spectral models presented in \cite{jerkstrand12}, we estimate a main-sequence mass for the progenitor star to be between $12 - 15$ M$_\odot$. After linear interpolation, we calculate a progenitor mass of $\approx 14$ M$_\odot$. This result is in agreement with other estimates presented in the literature: $13 \pm 1$ M$_\odot$ estimated via direct detection of the progenitor presented in \cite{xiang24}; 15 M$_\odot$ calculated via hydrodynamical modelling of the light curve presented by \cite{ertini25}; $\sim 15$M$_{\odot}$ derived by matching nebular-phase optical to mid-IR spectra between models and observations presented in \citep{dessart25}; $10 - 15$ M$_\odot$ from nebular-phase optical spectra compared with spectral models, and the [\ion{O}{1}] / [\ion{Ca}{2}]  flux ratio \cite{ferrari25}. 

\begin{acknowledgements}

We thank Dr. M. Barlow for the data sent via private communication. 
E.H. was financially supported by Becas-ANID scholarship \#21222163 and by ANID, Millennium Science Initiative, AIM23-0001. J.L.P. acknowledges support from ANID, Millennium Science Initiative, AIM23-0001. Based on observations obtained at the international Gemini Observatory under program GS-2024B-Q-417 (PI R. Cartier), a program of NSF’s NOIRLab, which is managed by the Association of Universities for Research in Astronomy (AURA) under a cooperative agreement with the National Science Foundation on behalf of the Gemini Observatory partnership: the National Science Foundation (United States), National Research Council (Canada), Agencia Nacional de Investigaci\'{o}n y Desarrollo (Chile), Ministerio de Ciencia, Tecnolog\'{i}a e Innovaci\'{o}n (Argentina), Minist\'{e}rio da Ci\^{e}ncia, Tecnologia, Inova\c{c}\~{o}es e Comunica\c{c}\~{o}es (Brazil), and Korea Astronomy and Space Science Institute (Republic of Korea). Based on observations obtained at the Southern Astrophysical Research (SOAR) telescope, which is a joint project of the Minist\'{e}rio da Ci\^{e}ncia, Tecnologia e Inova\c{c}\~{o}es (MCTI/LNA) do Brasil, the US National Science Foundation’s NOIRLab, the University of North Carolina at Chapel Hill (UNC), and Michigan State University (MSU). This paper includes data gathered with the 6.5 meter Magellan Telescopes located at Las Campanas Observatory, Chile under the program allocated by the Chilean Telescope Allocation Committee (CNTAC), no: CN2025A-89 and CN2025B-129 (P.I. E. Hueichapan).

\end{acknowledgements}

\bibliography{sample701}{}
\bibliographystyle{aasjournalv7}

\end{document}